\documentclass[runningheads]{llncs}
\usepackage{cite}
\usepackage{amsmath,amssymb,amsfonts}
\usepackage[normalem]{ulem}
\usepackage{algorithmic}
\usepackage{graphicx}
\usepackage{textcomp}
\usepackage{xcolor}
\usepackage{color}
\usepackage{upgreek}
\usepackage{multirow}

\newsavebox{\tablebox}
\usepackage{calc}
\usepackage{ulem}
\usepackage{adjustbox}
\usepackage{longtable}
\usepackage{wrapfig,lipsum,booktabs}

\def\BibTeX{{\rm B\kern-.05em{\sc i\kern-.025em b}\kern-.08em
    T\kern-.1667em\lower.7ex\hbox{E}\kern-.125emX}}

\usepackage{enumitem} %% allows specification of the tab size in enumerate, itemize, ... environments 

  {\begin{enumerate}%
    leftmargin = *
    \setlength{\itemsep}{0pt}%
    \setlength{\parskip}{0pt}}%
  {\end{enumerate}}

\usepackage[hyphens]{url}
\makeatother
\makeatletter

\let\llncssubparagraph\subparagraph
\let\subparagraph\paragraph
\usepackage[compact]{titlesec}
\let\subparagraph\llncssubparagraph

\begin{document}

\title{Automated Ransomware Behavior Analysis: Pattern Extraction and Early Detection
\thanks{\scriptsize{This manuscript has been authored by UT-Battelle, LLC, under contract DE-AC05-00OR22725 with the US Department of Energy (DOE). The US government retains and the publisher, by accepting the article for publication, acknowledges that the US government retains a nonexclusive, paid-up, irrevocable, worldwide license to publish or reproduce the published form of this manuscript, or allow others to do so, for US government purposes. DOE will provide public access to these results of federally sponsored research in accordance with the DOE Public Access Plan (http://energy.gov/downloads/doe-public-access-plan).”}.}}

 \author{ Qian Chen\inst{1}, 
     Sheikh Rabiul Islam\inst{2},
     Henry Haswell\inst{1}, 
     Robert A. Bridges\inst{3}}

 \authorrunning{Q. Chen et al.}
 
 \institute{Electrical and Computer Engineering Department, University of Texas at San Antonio, San Antonio TX\\
  \email{guenevereqian.chen@utsa.edu, henry.haswell@my.utsa.edu}
 \and
 Computer Science Department, Tennessee Technological University, Cookeville, TN \email {sislam42@students.tntech.edu}
 \and
 Computational Sciences \& Engineering Division, Oak Ridge National Laboratory, Oak Ridge, TN, \email{bridgesra@ornl.gov}
 }

\maketitle
\vspace{-5.8mm}
\begin{abstract}
Security operation centers (SOCs) typically use a variety of tools to collect  large volumes of host logs for detection and forensic of intrusions. 
Our experience, supported by recent user studies on SOC operators, indicates that operators spend ample time (e.g., hundreds of man hours) on investigations into logs seeking adversarial actions. 
Similarly, reconfiguration of tools to adapt detectors for future similar attacks is commonplace upon gaining novel insights (e.g., through internal investigation or shared indicators).  
 This paper presents an automated malware pattern-extraction and early detection tool, testing three machine learning approaches: \textit{TF-IDF} (term frequency–inverse document frequency), \textit{Fisher's LDA} (linear discriminant analysis) and \textit{ET} (extra trees/extremely randomized trees) that can (1) analyze freshly discovered malware samples in sandboxes and generate dynamic analysis reports (host logs);  (2) automatically extract the sequence of events induced by malware given a large volume of ambient (un-attacked) host logs, and the relatively few logs from  hosts that are infected with potentially polymorphic malware; (3) rank the most discriminating features (unique patterns) of malware and from the behavior learned detect malicious activity, and (4) allows operators to visualize the discriminating features and their correlations to facilitate malware forensic efforts. 
To validate the accuracy and efficiency of our tool, we design three experiments and test seven ransomware attacks (i.e., WannaCry, DBGer, Cerber, Defray, GandCrab, Locky, and nRansom).   
The experimental results show that \textit{TF-IDF} is the best of the three methods to identify discriminating features, and \textit{ET} is the most time-efficient and robust approach.
%that achieved an average 98\% to detect ransomware before data is encrypted. 
\end{abstract}

% \keywords{malware, ransomware, malware forensic, machine learning}

\section{Introduction}
Ransomware, a class of self-propagating malware,  uses encryption to hold victim's data and has experienced a 750\%  increase in frequency in 2018~\cite{Davis2018Ransomware}. Recently, the majority of these ransomware attacks target local governments and small business~\cite{DOBRAN2019Definitive}. For example, the 2018 SamSam ransomware hit the city of Atlanta, encrypted at least one third of users' applications, disrupted the city's vital services~\cite{STATESCOOP2019}, and resulted in \$17M of remediation to rebuild its computer network~\cite{Scmagazine2018}. 
Unlike large multinational businesses, small cities and businesses usually face stricter financial constraints than larger enterprises and struggle to establish or keep pace with cyber defensive technology and adversary/malware advancements. 
Consequently, they are less capable to defend against cyber threats. 
More generally, SOC's resource constraints and the shortage of cybersecurity talent~\cite{bridges2018information, goodall2004work, werlinger2010preparation} motivate us to develop an automated tools for SOCs.

Currently, manual investigation of logs is commonplace in SOCs and extremely tedious. 
E.g., our interaction with SOC operators revealed a 160 man-hour forensic effort to manually analyze a few CryptoWall 3.0 infected hosts' logs~\cite{chen2017automated} with the goal of (a) identifying the adversary/malware actions from user actions in their logs and (b) leveraging learned information to reconfigure tools for timely detection. 
This motivates our target use case---from SOC-collected logs from an attacked host (esp. a ransomware infection) and non-attack host logs, we seek to automated the (currently manual) process of identifying the attack's actions. In the ransomware case, this should be used to provides a pre-encryption ransomware detector. 
For testing in a controlled environment, we use ``artificial logs'', that is, logs obtained by running malware and ambient (emulated user) activities in a sandbox. 

Note that this mirrors classical dynamic analysis---(a) performing dynamic malware analysis to (b) extract indicators or signatures---and, hence, dynamic analysis is  a second use case.  
Malware analysis takes considerable time and requires an individual or a team with extensive  domain knowledge or reverse engineering expertise. Therefore, malware analysts usually collaborate across industry, university and government to analyze the ransomware attacks that caused disruptive global attacks (e.g., WannaCry). However,  the security community has insufficient resources to manually analyze less destructive attacks such as Defray, nRansom and certain versions of Gandcrab. Therefore,  manual analysis reports of such malware do not provide enough information for early detection~\cite{ Locky2016,Cerber-Gao, Locky2018, gandcrab,nransom, Defray2017,Defray2_2017,defray}.   Our approach, regardless of the  malware's real-world impacts and potential damages, efficiently help to automate tedious manual analysis by accurately extracting the most discriminating features from large amount of host logs and identifying malicious behavior induced by malware.

While our approach holds promise for more general malware and other attacks, we focus on ransomware. 
Note that upon the first infection identified in an enterprise, the logs from the affected host can be automatically turned into a detector via our tool. 
% Generalizing our target use case to include %, but more realistically, by SOCs using actual ambient host logs and those logs from an initial infection. Ideally the detector could be widely shared thereby extinguishing widespread, devastating ransomware attacks such as WannaCry at early onset. 
The tool applies three machine learning algorithms, (1) Term Frequency-Inverse Document Frequency (\textit{TF-IDF}), (2) Fisher's Linear Discriminant Analysis (\textit{Fisher's LDA}) and (3) Extra Trees/Extremely Randomized Trees (\textit{ET}) to (a) automatically identify discriminating features of an attack from system logs (generated by an automatic analysis system, namely, Cuckoo Sandbox~\cite{cuckoo}), and (b) detect future attacks from the same log streams.  
Using Cuckoo and set scripts for running ransomware and emulated user activity provides source data for experimentation with ground truth.  
We test the tool using infected system logs of seven disruptive ransomware attacks (i.e., WannaCry, DBGer, Cerber, Defray, GandCrab, Locky, and nRansom) and non-attack logs from emulated user activities, and present experiments varying log quality and quantity to test robustness. 
These system logs include  files,  folders, memory, network traffic, processes and API call activities. 

% Features generated from the system logs that have the largest TF-IDF scores and Fisher's LDA scores, or the features selected as non-leaf nodes that  have the smallest level of the decision tree model are ranked as top features.  
% The top-ranked features represent unique patterns of the ransomware attacks can provide high detection accuracy for ransomware early detection. 

\noindent \textbf{Contributions} of the pattern-extraction and early detection tool are
\vspace{-0.1cm}
\begin{enumerate}
    \item analyzing ransomware (esp. initial infection)  using Cuckoo Sandbox logs (more generally, ambient collected host logs)  and generating features from the host behavior reports. 
    \item extracting the sequence of events (features) induced by ransomware given  logs from (a few) hosts that are infected and (a potentially large amount of) ambient logs from presumably uninfected hosts;
    \item ranking the most discriminating features (unique patterns) of malware and identifying malicious activity before data is encrypted by the ransomware.
    \item creating graph visualizations of ET models to facilitate malware forensic efforts, and allowing operators to visualize discriminating features and their correlations.
\end{enumerate}

We compare outputs with ransomware intelligence reports, and validate that our tool is robust to variations of input data.  \textit{TF-IDF} is the best method to identify discriminating features, and \textit{ET} is the most time-efficient approach that achieves an average of 98\% accuracy rate to detect the seven ransomware. 
This work builds on preliminary results of our workshop paper~\cite{chen2017automated}, which only considered feature extraction, only used TF-IDF, and only tested with one ransomware.

% The paper is organized as follows. Section~\ref{sec:ransomware} introduces the seven ransomware attacks analyzed in this paper and compares our work with related works. Section~\ref{sec:approach} illustrates our research methodology.  Experiments and Results are discussed in Section~\ref{sec:exp}. We conclude the paper and discuss future work in Section~\ref{sec:con}.

\section{Background and Related Work} 
\label{sec:ransomware}
 
\noindent \textbf{Ransomware.} In contrast to the 2017 ransomware WannaCry that infected 300K machines across the globe, the majority of ransomware attacks in 2018 and 2019  have been targeting small businesses. These crypto-ransomware attacks usually use Windows API function calls to read, encrypt and delete files. Ransom messages  are displayed on the screen after the ransomware infecting the host. This paper selects and analyzes seven recently disruptive ransomware attacks.% that have struck from local government to global enterprises as follows.

\begin{enumerate}[leftmargin = *]
\vspace{-1mm}
    \item \textbf{WannaCry} \textit{(2017)}, a ransomware with historic world-wide effect, was launched on May 12, 2017~\cite{perlroth_2018}. The WannaCry dropper is a self-contained program consists of three components, an application encrypting and decrypting data;  an encryption key file; and a copy of Tor. WannaCry exploits vulnerabilities in Windows  Server Message Block (SMB) and propagates malicious code to infect other vulnerable machines on connected networks. %It reports that victims paidMore than \$143,000 worth of bitcoin to get their data back~\cite{Wannacry2017Hack}. 
 
    \item \textbf{DBGer} \textit{(2018)}, a new variant of the \textit{Satan} ransomware~\cite{Lemos2019Satan}, scans the victim local network for vulnerable computers with outdated SMB services.  DBGer incorporates a new open-source password-dumping utility, \textit{Mimikatz}, to store credential of vulnearble computers~\cite{DBGer}. The dropped Satan file is then executed to encrypt files of the infected computers with AES encryption algorithm. A text file \textit{\_How\_to\_decrypt\_files.txt} containing a note of demands from the attackers is displayed on victim's screen. 

    \item \textbf{Defray} \textit{(2017)}, a ransomware attack targets healthcare, education, manufacturing and technology industries~\cite{defray}. Defray propagates via phishing emails with an attached \textit{Word} document embedding an \textit{OLE} package object. Once the victim executes the OLE file, the Defray payload is dropped in the \textit{\%TMP\%} folder and disguises itself as an legitimate executable (e.g., \textit{taskmgr.exe} or \textit{explorer.exe}). Defray encrypts the file system but does not change file names or extensions. Finally,  it deletes volume shadow copies of the encrypted files~\cite{Defray2_2017}. Defray developers encourage victims to contact them and negotiate the payment to get the encrypted files back~\cite{Defray2017}. 

    \item \textbf{Locky} \textit{(2016, 2017)}  has more than 15 variants. It first appeared in February 2016 to infect Hollywood Presbyterian Medical Center in Los Angeles, California.  The ransomware attackers send millions of phishing emails containing attachments of malicious code that can be activated via \textit{Microsoft Word Macros}~\cite{Locky2018}. Locky encrypts data using RSA-2048 and AES-128 cipher that only the developers can decrypt data. In this research, we analyze the malicious behavior of a new variant of Locky ransomware called \textit{Asasin}, which encrypts and  renames the files with a  \textit{.asasin} extension.

    \item \textbf{Cerber} \textit{(2016-2018)} infected 150K Windows computers in July 2016 alone. Several Cerber variants appeared in the following two years have gained widespread distribution globally. Once the Cerber ransonware is deployed in the victim computer, it drops and runs an executable copy with a random name from the hidden folder created in \textit{\%APPDATA\%}. The ransomware also creates a link to the malware,  changes two Windows Registry keys, and encrypts files and databases offline with \textit{.cerber} extensions~\cite{Cerber2_2017,Cerber2016}.

    \item \textbf{GandCrab} \textit{(2018,2019)}, a Ransomware-as-a-Service (RaaS) attack has rapidly spread across the globe since  January, 2018. GandCrab RaaS online portal was finally shut down in June, 2019. During these 15 months, GandCrab creators regularly updated its code and  sold the malicious code, facilitating attackers without the knowledge to write their own ransomware~\cite{gandcrab2018}. Attackers then distribute GandCrab ransomware through compromised websites that are built with \textit{WordPress}. The newer versions of GandCrab use \textit{Salsa20} stream cipher to encrypt files offline instead of applying RSA-2048 encryption technique connecting to the C2 server~\cite{gandcrabv4.0}. GandCrab scans logical drives from \textit{A:} to \textit{Z:}, and encrypts files by appending a random Salsa20 key and a random initialization vector (IV) (8 bytes) to the contents of the file. The private key is encrypted in the registry using another Salsa20 key and the IV is encrypted with an RSA public key embedded in the malware. This new encryption method makes GandCrab a very strong ransomware, and the encrypted files can be decrypted by GandCrab creators only~\cite{GBV4.2}.
    
    \item \textbf{nRansom} \textit{(2017)} blocks the access to the infected computer rather than encrypting victim's data~\cite{nransom}. It demands ten nude photos of the victim instead of digital currency to unlock the computer. As recovery from nRansom is relatively easy, it is not a sophisticated malware but a ''test'' or a ''joke''.
\end{enumerate}
\vspace{-2mm}
\noindent \textbf{Ransomware Pattern Extraction and Detection Works.}  % Real-time ransomware detection and mitigation highly depend on the speed and accuracy of system logging. %logs mining. 
Homayoun et al. \cite{Homayoun2019Know} apply sequential pattern mining to find maximal frequency patterns (MSP) of malicious activities of four ransomware attacks. Unlike generating behavioral features directly from host logs, their approach summarizes activity using types of MSPs. % and six types of single step transition MSPs. %Atomic MSPs represent continuous events of the same type (e.g., file, registry and DLL), and single step transition MSPs represent a transition from one atomic MSP to another (e.g, FR means from file to registry). 
Using four machine learning classifiers, 
%(i.e., decision tree, random forest, bagging and MLP), 
the team found that atomic Registry MSPs  are the most important sequence of events to detect ransomware attacks with  99\% accuracy.

Verma et al. \cite{verma2018defining} embed host logs into a semantically meaningful metric space. 
The representation is used to build behavioral signatures  of ransomware  from host logs exhibiting pre-encryption detection, among other interesting use cases. 

Morato et al. introduces REDFISH~\cite{MORATO2018Ransomware},  a ransomware detection algorithm that identifies ransomware actions when it tries to encrypt shared files. % contained in shared network volumes from a Network Attached Storage.  
REDFISH is based on the analysis of passively monitored SMB traffic, and uses three parameters of traffic statistics to detect malicious activity. The authors use 19 different ransomware families to test REDFISH, which can detect malicious activity in less than 20 seconds.  REDFISH achieves a high detection rate but cannot detect ransomware before it starts to encrypt data. Our approach, discovering ransomware's pre-encryption footprint, promises a more accurate and in-time detection.

The Related Work section our preliminary work~\cite{chen2017automated} includes works published previously to those above. 
As the more general topic of  dynamic analysis is large and diverse,  a comprehensive survey is out of scope, but many exist, e.g. \cite{egele2012survey}.

%===

\section{Methodology}\label{methods}
\label{sec:approach}

The proposed approach requires a set of normal (presumably uninfected) system logs and at least one log stream containing ransomware behavior. 
% These are fed to three machine learning techniques to extract ransomware's behavior in the host logs. %the most discriminating features to discover the
In this study, 
%=====
%\begin{wrapfigure}{r}{0.6\linewidth}\centering
%\vspace{-0.5cm}
\begin{figure}[h]
\centering
\includegraphics[scale=.4]{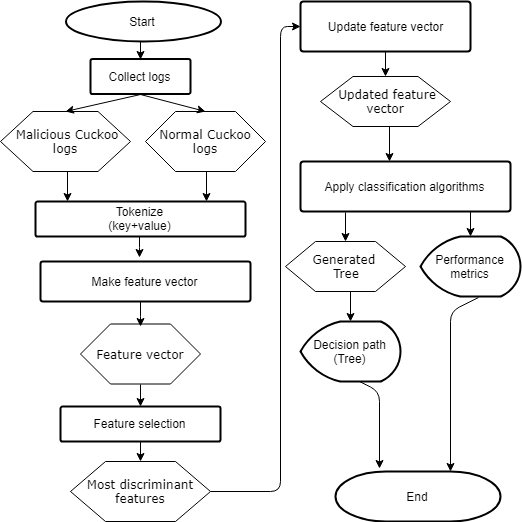}
\caption{ Flowchart of Research Methodology}
\label{fig:Flowchart}
%\vspace{0.1cm}
%\end{wrapfigure}
\end{figure}
% ======
the seven ransomware executables introduced in Section~\ref{sec:ransomware} are deployed inside a realistic but isolated environment with a sandbox tool, Cuckoo \cite{cuckoo}, for harvesting reproducible and shareable host logs. 
The Cuckoo host logs are dynamic analysis reports outlining  behavior (i.e., API calls, files, registry keys, mutexes), network traffic  and dropped files 
Meanwhile, Cuckoo also captures logs from scripted, emulated normal user activity such as reading and writing of executables,   deleting files, opening websites, watching YouTube videos, sending and receiving emails, searching flight tickets, and posting and deleting tweets on Twitter (see \cite{chen2017automated}).  
The normal user and the ransomware events/behavior in the raw host logs produced by Cuckoo are then converted to features, and  the three machine  
learning techniques are used to automatically obtain  the most discriminating features from normal and ransomware-including logs.  
%The most discriminating features of ransomware are the top-ranked features among all features of both normal and abnormal logs.  
Afterwards,  we discard the features that have little or no influence, and update  the feature vector to reduce the search space of  ET decision tree models.  The decision tree graphs are created to present the most discriminating features of ransomware attacks. See flowchart  in Figure~\ref{fig:Flowchart}.  %(i.e., the features that have zero TF-IDF and Fisher's LDA scores). 

\subsection{Feature Generation}
To build features we only use the \textit{enhanced} category and part of the \textit{behavior} category of Cuckoo-captured logging output. 
%in the   to the TF-IDF, LDA and ET measurable features.  
The details of the feature building can be found in our previous work~\cite{chen2017automated}. 
As malware often uses random names to create files, modules and folders, in this study,  we augment paths of specific files to emphasize their names only. For example,  \path{C:\\ Windows\\system32\\rsaenh.dll}  is converted to  a string ``c:..rsaenh.dll". Here,  ``.." is used as a wild-card to avoid generating duplicated features that represent similar host behavior.

\subsection{Discriminating Feature Extraction with Machine Learning}
TF-IDF, Fisher's LDA and ET are algorithms used in this research to automatically extract the most discriminating features of ransomware from host logs.

\textbf{TF-IDF}, was defined to identify the relative importance of a word in a particular document out of a collection of documents~\cite{salton1988term}. 
Our TF-IDF application follow our previous work for accurate comparison. 
Given two sets of documents %(in our case one with logs containing ransomware activities, other  some containing normal activities) 
let $f(t,d)$ denote the frequency of term $t$ in document $d$, and $N$ the size of the corpus. The TF-IDF weight is the product of the Term Frequency,  $\mbox{tf}(t,d)=f_{t,d}/\sum_{t'\in d} f_{t',d}$ 
(giving the likelihood of $t$ in $d$) and the Inverse Document Frequency,
$\mbox{idf}(t,D) = \log [ {N}/(1+|\{d\in D: t\in d\}|) ]$ (giving the Shannon's information of the document containing  $t$). 
Intuitively, given a document, those terms that are uncommonly high frequency in that document are the only terms receive high scores. 
%Our application is to consider a log stream from a host as a document. 
We use log streams from infected hosts as one set of documents and a set of normal log streams as the other to apply TF-IDF; hence, highly ranked features  occur often in  (and are guaranteed to occur at least once in) the ``infected''  document, but infrequently anywhere else~\cite{chen2017automated}. 

\textbf{Fisher's LDA} is a supervised learning classification algorithm that operates by projecting the input feature vectors to a line that (roughly speaking) maximizes the separation between the two classes~\cite{welling2005fisher}. For our application we consider a binary classification where one class  ($C_1$) is comprised of the feature vectors $\{x_i\}_i \subseteq \mathbb{R}^m$ representing host logs that included ransomware,   and the second class ($C_2$) are those vectors of ambient logs. 
We use this classifier for identifying the discriminating features between the classes. 
Consider the set $\{v^tx_i : x_i \in C_1 \cup C_2\}\subset \mathbb{R}$, which is the projection of all feature vectors to a line in $\mathbb{R}^m$ defined by unit vector $v$. 
Fisher's LDA identifies the unit vector $v$ that maximizes $S(v):= [v^t (\mu_1 -\mu_2)]^2 / [v^t(\Sigma_1 + \Sigma_2)v]$ with $\mu_j, \Sigma_j$ the mean and covariance of $C_j, j = 1,2$, respectively. $S(v)$ is the squared difference of the projected classes' means divided by the sum of the projected classes' variances. 
It is an exercise in linear algebra to see the optimal $v \propto (\Sigma_1 + \Sigma_2)^{-1}(\mu_1 - \mu_2).$ 
Geometrically, $v$ can be thought of as a unit vector pointing from $C_1$ to $C_2$;  hence, ranking the components of $v$ by absolute values sorts the features that most discriminate the ransomware and normal activity. 
% Let $\mu_k$ be the means of two classes $C_k$ where $k = 1$ or$ 2$ and $S_k$ represents separate class scatter matrix for each class, where $S_k=\sum_{x_i\in c_k} (x_i-\mu_k)(x_i-\mu_k)^t$. 
% $S_w$ is  the class scatter matrix, and $S_w$= $S_1$ + $S_2$. Thus, the optimal line direction $v$=$S_{w}^{-1} (\mu_1-\mu_2)$, and the 1-D projection of $x_i$ (i.e.,$v^tx_i$) can be obtained following the formula $Y_k$=$v^tc_k^{t}$. 

\textbf{Extremely Randomized Trees} (ET) is a tree-based ensemble algorithm for supervised classification and regression. ``It consists of randomizing strongly both attribute and cut point choice while splitting the tree node" \cite{geurts2006extremely}. 
In  the extreme case, the algorithm provides ``totally randomized trees whose structures are independent of the output values of the learning sample"~\cite{geurts2006extremely, islam2018credit}. The randomization introduces increased bias and variance of individual trees. However, the effect on variance can be ignored when the results are averaged over a large ensemble of trees. This approach is tolerant with respect to over-smoothed (biased) class probability estimates~\cite{geurts2006extremely}. See the cited works for details.

\section{Experimental Results}
\label{sec:exp}
% We present three experiments to test the pattern extraction and early detection tool with seven ransomware attacks. \\

\noindent \textbf{Experiment One: Extracting Discriminating Features from Host Logs}. This experiment applies the machine learning approaches to extract the most discriminating features/behavior of each ransomware attack. % from a large amount of normal behavior logs. 
In addition to obtaining a Cuckoo analysis report (raw behavior log) for each ransomware sample, Python scripts imitating  various users’ normal activities (such as reading, writing and deleting files, opening
websites, etc.) are submitted to the Cuckoo sandbox to generate a large volume of normal reports. 
\begin{table}[]
\vspace{-1cm}
\centering
\caption{The Most Discriminating Features of the Seven Ransomware Attacks}
\label{tab:actions} 
\begin{adjustbox}{width=1\textwidth}
\begin{tabular}{|c|l|l|c|c|c|}

\hline
\multirow{2}{*}{\#} & \multicolumn{1}{c|}{\multirow{2}{*}{Pre-Encryption Pattern}} & 
\multicolumn{1}{c|}{\multirow{2}{*}{Feature}} & 
\multicolumn{3}{c|}{Rank}   \\ \cline{4-6}  &     &   
& TF-IDF                 & LDA                       & ET                                                  \\ \hline
\parbox[t]{2mm}{\multirow{7}{*}{\rotatebox[origin=c]{90}{1. WannaCry}}}   & 1. Import CryptoAPI from advpi32.dll   & data\_file+`advapi32.dll'+event+`load'+object+`library'                                                                                                                           & \multicolumn{1}{c|}{3} & \multicolumn{1}{c|}{294} & 6                                                  \\ \cline{2-6} 
                            & 2. Unzips itself to .wrny files        & *.wnry                                                                                                                                                                            & \multicolumn{1}{c|}{1} & 176                      & 1                                                   \\ \cline{2-6} 
                            %& 3. Generates machine-unique identifier & \begin{tabular}[c]{@{}l@{}} thsgvkvtwaipdcd971\end{tabular}                                                           &    NA                    &      NA                     &                NA                                     \\ \cline{2-6} 
                            & 3.  \begin{tabular}[c]{@{}l@{}} 
        Creates a registry, \path{HKEY_LOCAL_MACHINE\}\\ \path{Software\WanaCrypt0r\wd}
        \end{tabular}                                        & \begin{tabular}[c]{@{}l@{}}api+`regcreatekeyexw'+arguments\_1\_value+`33554432'\\+category+`registry' \textit{(subkey=``Software\textbackslash{}\textbackslash{}WanaCrypt0r")}\end{tabular} & 6                     & 177                      & \begin{tabular}[c]{@{}l@{}} NA \end{tabular} \\ \cline{2-6} 
                            & 4. \begin{tabular}[c]{@{}l@{}}  
        Run `attrib +h', to set the current directory \\as a hidden folder
        \end{tabular}                                           & data\_file+`attrib +h .'+event+`execute'+object+`file'                                                                                                                            &  6                     & 298                     & 11                                                  \\ \cline{2-6} 
                            & 5.       \begin{tabular}[c]{@{}l@{}}
            Run `icacls . /grant Everyone:F /T /C /Q' to \\grant user permissions to the current directory
        \end{tabular}                                         & data\_file+`icacls . ..q'+event+`execute'+object+`file'                                                                                                                           & 6                      & 298                     & 11                                                  \\ \cline{2-6} 
                            & 6.  \begin{tabular}[c]{@{}l@{}}
            Import public and private RSA AES \\
            keys (000.pky, 000.eky)  from t.wrny\\ 
        \end{tabular}                                      & data\_file+`c:..00000000.pky'+event+`write'+object+`file'                                                                                                                         & 6                      & 298                     & 11                                                  \\ \hline \hline

  %=======DBGER
  \parbox[t]{2mm}{\multirow{3}{*}{\rotatebox[origin=c]{90}{2. DBGer}}}   & \begin{tabular}[c]{@{}l@{}}1. Drop ExternalBlue files at\\`C:\textbackslash{}Users\textbackslash{}All Users\textbackslash{}' \end{tabular}& \begin{tabular}[c]{@{}l@{}}data\_file+`c:..users'+event+`create'+object+`dir',\\ data\_file+`c:..allusers'+event+`create'+object+`dir',\\ data\_file+`c:..blue.exe'+event+`write'+object+`file',\\ \textbf{... 22 various dropped file features...}\\ data\_file+`c:.. satan.exe’ +event+`write'+object+`file',\\ data\_file+`c:..mmkt.exe’ +event+`write'+object+`file'\end{tabular} & 9                      & 125                       & \begin{tabular}[c]{@{}l@{}}11,\\ 12\end{tabular}    \\ \cline{2-6} 
                                   
   & \begin{tabular}[c]{@{}l@{}}2. Drop satan.exe on C drive\\ and execute the file for encryption\end{tabular}                                         & \begin{tabular}[c]{@{}l@{}}data\_file+`c:..satan.exe'+event+`write'+object+`file',\\ data\_file+`c:..satan.exe'+event+`execute'+object+`file'\end{tabular}      
   & 9                      & 125                       & \begin{tabular}[c]{@{}l@{}}11,\\ 12\end{tabular}    \\ \cline{2-6} 
   & 3. Drop ``KSession" file at \%Temp\% & data\_file+`c:..ksession'+event+`write'+object+`file'                & 9                      & 125                       & 11                                                  \\ \hline
    \hline
%=====Defray

%\parbox[t]{2mm}{\multirow{3}{*}{\rotatebox[origin=c]{90}{3. Defray}}} 
\multirow{3}{*}[-1ex]{\rotatebox{90}{3. Defray}}& \multicolumn{1}{l|}{1. Import/Load Microsoft OLE from ``ole32.dll"}                                                                                                                             & \multicolumn{1}{l|}{data\_file+`ole32.dll'+event+`load'+object+`library'}                                                                                                                                                                                                                                                                                                                       & \multicolumn{1}{c|}{9}      & \multicolumn{1}{c|}{10}    & \multicolumn{1}{c|}{9}                                                   \\ \cline{2-6} 
\multicolumn{1}{|l|}{}                            & \multicolumn{1}{l|}{2. Drop and execute ``explorer.exe"}                                                                                                              & \multicolumn{1}{l|}{data\_file+`explorer.exe'+event+`load'+object+`library'}                                                                                                                                                                                                                                                                                                                    & 17     &93   & NA                                           \\ \cline{2-6} 
             & \begin{tabular}[c]{@{}l@{}}3. Call ShellExecute to run as more privileged\\user to disable startup recovery and delete\\volume shadow copies\end{tabular}               & \begin{tabular}[c]{@{}l@{}}data\_file+`c:..-hibernate-timeout-dc0'+event+`execute'\\+object+`file'          \end{tabular}   & 17     & 121   & NA    \\  \hline          \hline
   
  %=====Locky
  %\parbox[t]{2mm}{\multirow{5}{*}{}{\rotatebox[origin=c]{90}{4.Locky}}}     
  \multirow{5}{*}[-4ex]{\rotatebox{90} {4.Locky}}& \begin{tabular}[c]{@{}l@{}} 1. Read and write `PIPE\textbackslash{}\textbackslash{}wkssvc' and\\ `PIPE\textbackslash{}lsarpc'  \end{tabular}                                                                     & \begin{tabular}[c]{@{}l@{}}data\_file+`pipe..wkssvc'+event+`write'(`read')+object+`file',\\ data\_file+`pipe..lsarpc'+event+`write' (`read')+object+`file'\end{tabular}                                                                                                                                                                                                    & \begin{tabular}[c]{@{}l@{}}2 \\7\end{tabular} & \begin{tabular}[c]{@{}l@{}}72 \\408\end{tabular} & \begin{tabular}[c]{@{}l@{}}2\\ 7\end{tabular}  \\ \cline{2-6} 
                            & 2. Read network provider name                                                                                                                                              & \begin{tabular}[c]{@{}l@{}}data\_regkey+`hkey\_local\_machine..\\networkprovidername'+event+`read'+object+`registry'\end{tabular}                                                                                                                                                                                                                                         & 3                                                & 186                                                 & 3                                                   \\ \cline{2-6} 
                            & 3. Read the path to the network provider .dll file                                                                                                                         & \begin{tabular}[c]{@{}l@{}}data\_regkey+`hkey\_local\_machine..\\systworkproviderproviderpath'+event+`read'+object+`registry'\end{tabular}                                                                                                                                & 4                                                & 171                                                 & 4                                                   \\ \cline{2-6} 
                            & 4. Load the network provider `ntlanman.dll" file                                                                                                                      & data\_file+`c:..ntlanman.dll'+event+`load'+object+`library'                                                                                                                       & 4                                                & 130                                                 & 4                                                   \\ \cline{2-6} 
                            & 5. Obtain the name of the Security Identifier                                                                                                                              & \begin{tabular}[c]{@{}l@{}}data\_regkey+`hkey\_users..s-1-5-21-1966058-1343024091\\-1003name'+event+`read'+object+`registry'\end{tabular}                                             & 5            & 408     & 5                                                   \\ \hline
                            \hline
%========Cerber
\multirow{4}{*}[-10ex]{\rotatebox{90} {5.Cerber}}          & \begin{tabular}[c]{@{}l@{}}1. Create two .tmp files under a random folder \\ in \%APPData\%\end{tabular} & \begin{tabular}[c]{@{}l@{}}a. data\_file+`c:..b51826f3'+event+`create'+object+`dir'\\ b. data\_file+`c:..4e89.tmp'+event+`write'+object+`file'\\ c. data\_file+`c:..5572.tmp'+event+`write'+object+`file'\end{tabular}         & 10                                                      & \begin{tabular}[c]{@{}l@{}}a.105\\ b.230\\c.230\end{tabular}        & \begin{tabular}[c]{@{}l@{}}a.10\\b.10\\ c.11\end{tabular}   \\ \cline{2-6} 
                            & 2. Find users profiles and read the profiles                                                             & \begin{tabular}[c]{@{}l@{}}a.data\_regkey+`hkey\_local\_machine..\\softilelistprofilesdirectory'+event+`read'+object+`registry'\\ b.data\_regkey+`hkey\_local\_machine..\\softlelistdefaultuserprofile'+event+`read'+object+`registry'\\ c. data\_regkey+`hkey\_local\_machine..\\softs-1-5-18profileimagepath' +event+`read'+object+`registry'\\ \textbf{...omit SID 1-5-19$\sim$1-5-20...}\\ d. data\_regkey+`hkey\_local\_machine..\\soft091-1003profileimagepath +event+'read'+object+`registry'\end{tabular} & \begin{tabular}[c]{@{}l@{}}a.5\\b.5\\c.7\\d.7\end{tabular}   & \begin{tabular}[c]{@{}l@{}}a.111\\b.111\\ c.150\\d.150\end{tabular}      & \begin{tabular}[c]{@{}l@{}}a.5\\b.5\\ c.7\\d.7\end{tabular}    \\ \cline{2-6} 
                            & 3. Read and load ``rsaenh.dll"                                                                           & \begin{tabular}[c]{@{}l@{}}a. data\_regkey+`hkey\_local\_machine..\\ softaphic providerimage path'+event+`read'+object+`registry'\\ b. data\_file+`c:..rsaenh.dll'+event+`read'+object+`file'\\ c. data\_file+`c:..rsaenh.dll'+event+`load'+object+`file'\end{tabular}                                                                                                                             & \begin{tabular}[c]{@{}l@{}}a.3\\ b.1\\ c.6\end{tabular} & \begin{tabular}[c]{@{}l@{}}a.79\\ b.15\\ c.119\end{tabular} & \begin{tabular}[c]{@{}l@{}}a.3\\ b.1\\ c.6\end{tabular} \\ \cline{2-6} 
                            & 4. Obtain Machine GUID from registry                                                                     & \begin{tabular}[c]{@{}l@{}}data\_regkey+`hkey\_local\_machine..\\ cryptographymachineguid'+event+`read'+object+`registry'\end{tabular}                                  & 2        &69           & 2                                                       \\ 
                            \hline
                            \hline

%=======GandCrab

\multirow{2}{*}[-1ex]{\rotatebox{90} {6.Gandcrab}}          & \begin{tabular}[c]{@{}l@{}}1. Scan and collect information\\     a. computer name \\    b. session manager name\\    c. domain name\\    d. processor type\end{tabular} & \begin{tabular}[c]{@{}l@{}}a.data\_regkey+`hkey\_local\_machine..systcomputername\\computername'+event+`read'+object+`registry'\\ b.data\_regkey+`hkey\_local\_machine..sessionmanagername'\\ +event+`read'+object+`registry'\\ c. data\_regkey+`hkey\_local\_machine..parametersdomain'\\ +event+`read'+object+`registry'\\ d.1 data\_regkey+`hkey\_local\_machine..0processornamestring'\\ +event+`read'+object+`registry'\\ d.2 data\_regkey+`hkey\_local\_machine..0identifier'\\ +event+`read'+object+`registry'\\ d.3 data\_regkey+`hkey\_local\_machine..\\systgersafeprocesssearchmode'+event+`read'+object+`registry'\end{tabular} & \begin{tabular}[c]{@{}l@{}}a.1\\ b.6\\ c.7\\ d.7\end{tabular} & \begin{tabular}[c]{@{}l@{}}a.276\\ b.430\\ c.431\\ d.431\end{tabular} & \begin{tabular}[c]{@{}l@{}}a.1\\ b.8\\ c.10\\ d.9\end{tabular} \\ \cline{2-6} 
& \begin{tabular}[c]{@{}l@{}}2. Copy the ransomware .exe file\\ to \%APPDATA\%/Microsoft\\ and add an entry to RunOnce key\end{tabular}                                                 & \begin{tabular}[c]{@{}l@{}}a. data\_file+`c:..lrcjty.exe'+event+`write'+object+`file'\\ b. data\_content+`..x00'+data\_object+`none'+data\_regkey+\\ `hkey\_current\_user..runonceoopmhnlocoz'\\ +event+`write'+object+`registry'\end{tabular}          & 7    & 431  & \begin{tabular}[c]{@{}l@{}}a. 9\\ b.10\end{tabular}  \\ \hline
\hline
%==========nRansomware======
\multirow{4}{*}[-5ex]{\rotatebox{90} {7.nRansomware}}          & \begin{tabular}[c]{@{}l@{}}1. Create temprary directory in  \textbackslash{}\%TEMP\%\textbackslash{}1.tmp\textbackslash{}\\tools\textbackslash{} \end{tabular} & \begin{tabular}[c]{@{}l@{}}data\_file+`c:..tools'+event+\textbf{`create'+object+`dir'} \end{tabular}         & 5                                                      & \begin{tabular}[c]{@{}l@{}} 32 \end{tabular}        & \begin{tabular}[c]{@{}l@{}} 5 \end{tabular}   \\ \cline{2-6} 
                            & 
                            \begin{tabular}[c]{@{}l@{}}2. Download and write following files: \\  a. an executable (i.e., nransom.exe) \\ 
                            b. a media control file (i.e.,interop.wmplib.dll) \\
                            c. a audio file (i.e., your-mom-gay.mp3)
                            \end{tabular}
                            & \begin{tabular}[c]{@{}l@{}}a.data\_file+`c:..\textbf{nransom.exe'+event+\textit{`write'}}+object+`file'\\ b.data\_file+`c:..\textbf{interop.wmplib.dll'+event+`\textit{write'}}+object+`file'\\ 
                            c.data\_file+`c:..\textbf{your-mom-gay.mp3'+event+\textit{`write'}}+object+`file' \end{tabular} & \begin{tabular}[c]{@{}l@{}}a.4\\b.4\\c.4\end{tabular}   & \begin{tabular}[c]{@{}l@{}}a.23\\b.23\\ c.23\end{tabular}      & \begin{tabular}[c]{@{}l@{}}a.4\\b.4\\c.4\end{tabular}    \\ \cline{2-6} 
                            & \begin{tabular}[c]{@{}l@{}}3. Execute the executable  (i.e., nransom.exe)  using \\ command prompt (i.e.,cmd.exe) that lock the screen
                            \end{tabular}                                                                   & \begin{tabular}[c]{@{}l@{}}a.  data\_file+\textbf{`nransom.exe'+event+\textit{`execute'}}+object+`file'\\ b. data\_file+`c:..\textbf{cmd'+event+\textit{`execute'}}+object+`file' \end{tabular}                                                                                                                             & \begin{tabular}[c]{@{}l@{}}a.6\\ b.6\end{tabular} & \begin{tabular}[c]{@{}l@{}}a.60\\ b.60\\ \end{tabular} & \begin{tabular}[c]{@{}l@{}}a.7\\ b.6\end{tabular} \\ \cline{2-6} 
                            &                                   \begin{tabular}[c]{@{}l@{}} 4. Play the looped song using the  downloaded \\ audio file (i.e., your-mom-gay.mp3)   \end{tabular}                                 & \begin{tabular}[c]{@{}l@{}}data\_file+`c:..\textbf{your-mom-gay.mp3'+event+\textit{`read'}} \\ +object+`file'\end{tabular}                                  & 5        & 32           & 5                                                       \\  \cline{2-6}
                            &\begin{tabular}[c]{@{}l@{}} 5. Delete the temporary folders with \\ the downloaded files    \end{tabular}                                 & \begin{tabular}[c]{@{}l@{}} data\_file+`c:..\textbf{1.tmp'+event+`\textit{delete'}}+object+`dir' \\ +object+`file'\end{tabular}                                  & 6        & 60           & 6                                                       \\ 
                            \hline
                            \hline

\end{tabular}
\end{adjustbox}
\vspace{-1cm}
\end{table}

Table~\ref{tab:actions} illustrates the most discriminating features of the seven ransomware attacks.  The first column of the table (\textit{\#}) lists the name of seven ransomware. The second column (\textit{Pattern}) presents the pre-encryption patterns (activities) of each ransomware attack obtained from the detailed ransomware technical (static) analysis produced by cybersecurity companies (e.g., FireEye~\cite{wannacry}), security help websites (e.g., Bleeping Computer~\cite{cimpanu_2018,locky}) and malware research teams (e.g., The Cylance Threat Research~\cite{defray}).  The third column (\textit{Feature}) presents the features extracted from the host logs using the proposed approaches that match the unique patterns of rasomware attacks. 
The last column (\textit{Rank}) lists the TF-IDF, Fisher's LDA and ET rankings of the features that represent the unique patterns of the seven ransomware attacks.   
The features that have the largest TF-IDF and Fisher's LDA scores, or the non-leaf nodes (features) of the Extremely Randomized Trees that have smallest levels, are top-ranked discriminating features. 
For the ET algorithm, the features that are at the top of the tree contribute more to correctly classifying a larger portion of input logs. 
E.g., a feature with rank = 1 is one of the most indicative feature of the malware according to that algorithm.  
Ties are possible as the scores may be the same between multiple features.
We use the rankings of these features to evaluate the efficiency %and accuracy 
of the proposed three machine learning methods. 
The methods that provide higher rankings of the selected features are more efficient than the approaches that yield a lower rank of the same feature. 

% The relative rank (i.e., level) of a feature (or non-leaf node in the decision tree) is reported. %measures the ity (i.e., a measure of incorrectly labeling). In other words, 

We set a large \textit{class\_weight} parameter  for the target class in  \textit{ExtraTreesClassifier} of Python's \textit{Scikit-Learn} library to make the ET classifier biased to learn the pattern of malicious logs more meticulously. Therefore, some features representing the ransomware patterns are not selected as the nodes to compose the tree.  In this scenario, we use ``NA" to present the rankings of the feature that are not nodes in the tree. Details are elaborated by ransomware: 
\vspace{-0.2cm}
\begin{enumerate}[leftmargin = *]
    \item\textbf{WannaCry:} The six patterns of WannaCry before the attack encrypting  data are presented in Table~\ref{tab:actions}. All of these patterns can find WannaCry-generated features from the host logs. A total of $1,207$ unique features have been extracted from host logs containing both normal and abnormal behavior, while only a small portion are resulting from WannaCry actions. The experimental results indicate that TF-IDF is better than the other two methods for identifying WannaCry's behaviors.  The rankings generated by the ET classifier are slightly lower than the TF-IDF's. However, ET is more time efficient for extracting the most discriminating features from large volume of host logs, which requires only $215$ features (nodes) to make decisions (i.e., WannaCry or Normal).  Therefore, the results suggest using TF-IDF to analyze the few infected hosts’ logs in an attempt to produce shareable threat intelligence reports and using the ET algorithm to obtain pre-encryption detection capabilities. 
    
    This experiment also illustrates that the top-ranked features generated by Fisher's LDA are quite different from the other two techniques. Most of the top-ranked features are normal activities. Features representing WannaCry's patterns are listed as low as \#200. Additionally, we notice that the loading and reading events of the \textit{rsaenh.dll} module are ranked highly (i.e., \#2 and \#4  for TF-IDF and \#3 and \#8 for ET). The module implements the Microsoft enhanced cryptographic service provider for WannaCry to encrypt the victim's data with 128-bit RSA encryption. These two top ranked features are not listed in our table, as they are not  discriminating features to identify WannaCry attacks from other crypto-ransomware attacks. 
    
    \item \textbf{DBGer:} The three unique patterns of DBGer ransomware reported by \cite{DBGer-Tech} are presented in Table~\ref{tab:actions}. \textit{dbger.exe}, the mother file of DBGer, first creates the \path{C:\Users\AllUsers} folder, drops \textit{EternalBlue} and \textit{Mimikatz} executables in the new folder, and then saves \textit{satan.exe} into the \textit{C} drive. A file named \textit{KSession} is dropped to \path{C:\Windows\Temp\ } for storing the host ID. TF-IDF and Fisher's LDA rank $1,104$ features generated from normal and DBGer Cuckoo reports. The ET classifier builds the decision tree using $216$ of the $1104$ features. The three DBGer features are ranked highly.  TF-IDF yields a highest ranking of the three features, which is better than the other two methods. ET is more time efficient. However, there are many features ranked higher than the ranking of the three  features, but they are normal activity. E.g.,   dynamic link library (DLL) files \textit{kernel32.dll} and \textit{advapi.dll} are on the top of the three rankings, but are not discriminating features for DBGer.   %Additionally, one of the top-ranked features contains \textit{dbger.exe}. However,  this feature is not As malware usually changes its executable to a random name, this feature . 

    \item \textbf{Defray:} The three unique patterns of Defray are loading the \textit{ole32.dll} file, dropping and executing the ransomware executable file \textit{explorer.exe},  and executing a shell command. The three machine algorithms rank the first feature ``loading the ole32.dll file" \#9 among the total $1,243$ features. As Defray's executable file is disguised as a Windows Internet Explorer, all of the three methods struggle to distinguish it from the normal activities. The second feature therefore is not selected to build the ET model, and its TF-IDF and Fisher's LDA weights are much lower than the first feature's. The three machine learning approaches rank another three features (as shown in Table~\ref{tab:missed}) highest among the $1243$ features. These features represent unique malicious activities performed by Defray, thus, they are discriminating features to distinguish Defray from other ransomware.  However, none of these three patterns are discussed in Defray manual analysis reports~\cite{Defray2017,defray,Defray2_2017}.

    \item \textbf{Locky:} We execute \textit{Asasin Locky},  a 2017 variant of Locky ransomware in the Cuckoo sandbox, collect and analyze its behavior using our tool.  The static analysis reports~\cite{Locky2016, Locky2018} indicate that after being deployed, Locky's executable file disappears. Its dropped copy \textit{svchost.exe} is executed from the \textit{\%TEMP\%} folder. However, our tool generates features from the behavior logs and presents that Asasin Locky does not drop the executable file. Instead, the attack modifies the workstation services \path{\PIPE\wkssvc} launched by the \textit{svchost.exe} process. As a member of the Cryptowall family, Asasin Locky also modifies \path{PIPE\lsarpc}, a file communicates with the Local Security Authority subsystem~\cite{MONIKA2016Experimental}. The attack then reads network provider name and the path to the Network Provider DLL file from registry by loading the network provider \textit{ntlanman.dll}. Registry is retrieved by Asasin Locky to obtain the name of the Security Identifier.  TF-IDF and ET provides the same and higher rankings for these five features from a total $1,047$ normal and ransomware features. These two methods both rank \textit{rsaenh.dll} as the top feature; however, this feature is not a unique pattern for Asasin Locky.  

    \item \textbf{Cerber:} This ransomware copies itself as \textit{cerber.exe} to the hidden \textit{\%APPDATA\%} folder, creates a directory with a random name, and drops two \textit{.tmp} files~\cite{Cerber-Gao}. Cerber also escalates its privilege to admin level and reads profiles from the users' profile image paths. Afterwards, Cerber finds the image path of \textit{rsaenh.dll}, reads and loads the DLL file to encrypt data. Cerber obtains the Machine GUID (globally unique identifier) and uses its fourth part as the encrypted files' extension. The Cerber sample tested has an extension of \textit{93ff}. The three methods rank the total $1,137$ features. ET selects $145$ features to composes the decision tree. TF-IDF and ET provides similar and higher rankings of the discriminating features than Fisher's LDA's. 
    
    \item \textbf{GandCrab:} This experiment uses Gandcrab V2.3.1, a variant that  scans the victim machine and collects information of user name, domain name, computer name, session manager name and processor type~\cite{gandcrab}. The execution is terminated if the ransomware finds the system language is Russian or the victim machine installed specific anti-virus (AV) software. Otherwise, it copies the executable file into \path{\%APPDATA\%/Microsoft} and adds an entry of the copied executable file path to the \textit{RunOnce} key as a one-time persistence mechanism. GandCrab then decrypts the ransom notes and generate RSA keys for encryption. After encrypting data, the malware uses Windows' NSLOOKUP tool  to (1) find IP address of the GandCrab's C2 (command and control) server; and (2) communicate with the C2 server (i.e., sending information collected from the victim’s machines  to the C2 server and/or receiving commands from the C2 server). Table~\ref{tab:actions} presents two unique pre-encryption patterns of GandCrab V2.3.1. TF-IDF and ET rank them highly among $1,017$  features. The rankings of these features are much lower by Fisher's LDA.

    \item \textbf{nRansom:} This attack first creates a subfolder in \textit{\%TEMP\%} with a random name ended with~\textit{.tmp}. In our experiment, the subfolder is named \textit{1.tmp}.  nRansom  drops an executable file (i.e., \textit{nransom.exe}) and two Windows Media Player control library files (i.e., \textit{Interop.WMPLib.dll} and \textit{AxInterop.WMPLib.dll}) in \textit{1.tmp}. An audio file \textit{your-mom-gay.mp3} is dropped in \path{1.tmp\Tools}. Then \textit{nransom.exe} is executed through the command prompt \textit{cmd.exe}. After locking the victim's computer screen, nRansom plays a looped song from the dropped mp3 file, and deletes the subfolders and dropped files. TF-IDF and ET both rank the five discriminating features of nRansom highly among $1046$ features.  $55$ features are used for composing ET. 
 \end{enumerate}
 
%======
\begin{table}[!ht]
%\vspace{-0.2cm}
\centering
\caption{Static Analysis Missed Unique Patterns and Their Behavioral Features}
\label{tab:missed} 
\centering
  \includegraphics[width=1\textwidth]{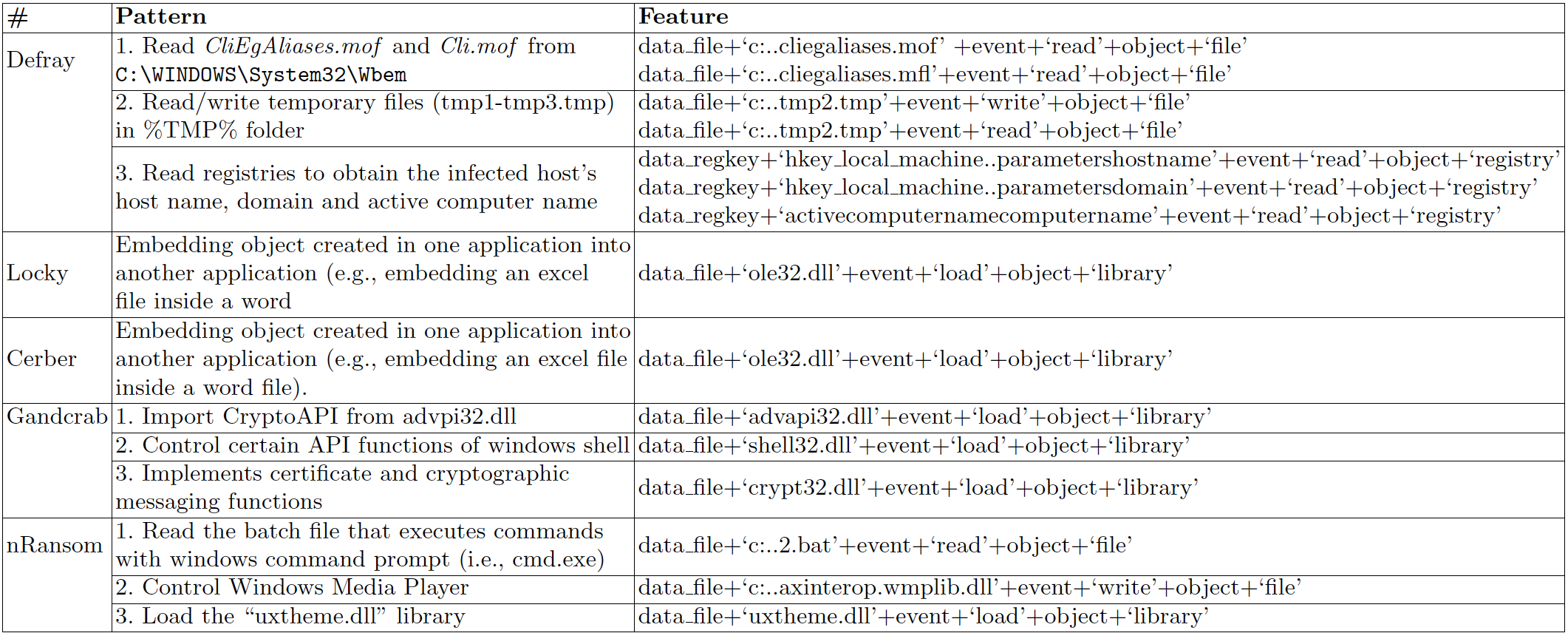}
\vspace{-0.2cm}
\end{table}
%======== 
\noindent \textbf{Ransomware Unique Patterns Missed from Manual Analysis.}\\
As discussed above, besides the patterns obtained from Defray's threat intelligence reports, the three features shown in Table~\ref{tab:missed} are also unique behavior to distinguish Defray attacks. From the dynamic analysis provided by our methodology, we also found that many ransomware attacks have similar patterns. For example, Defray, Locky and Cerber all conduct an event to load the \textit{ole32.dll} file. 
However, neither Locky nor Cerber's static analysis have mentioned this pattern. 
Similarly,  manual analysis of GandCrab does not discuss the malware sample has imported CryptoAPI from \textit{advapi32.dll}, which is also a discriminating feature of WannaCry attacks. 
Thus, our tool provides automated---more efficient and without reliance on security experts---and better quality  malware behavior analysis. 

\noindent \textbf{Experiment Two: Ransomware Feature Rankings with Varying Normal Activities}. 
This experiment aims to validate that the rankings of the seven ransomware discriminating features are not influenced by varying the number of normal logs. To validate the hypothesis, we calculate the 
TF-IDF, Fisher's LDA and ET weights of the ransomware features  in the following three scenarios. 
%=====
\begin{itemize}
\vspace{-1mm}
  \item Case 1 (C1): Using Experiment One's normal logs as the baseline.  
  \item Case 2 (C2): Adding 30\% additional new normal host logs into training data.
  \item Case 3 (C3): Adding 60\% more new normal host logs into training data.
  \vspace{-3mm}
\end{itemize}
%====

\begin{table}[!ht]
\centering
%\vspace{-1cm}
\caption{WannaCry Discriminating Feature Ranking with Varying Normal Data}
  \label{table:experiment2}
 \centering
  \includegraphics[width=1\textwidth]{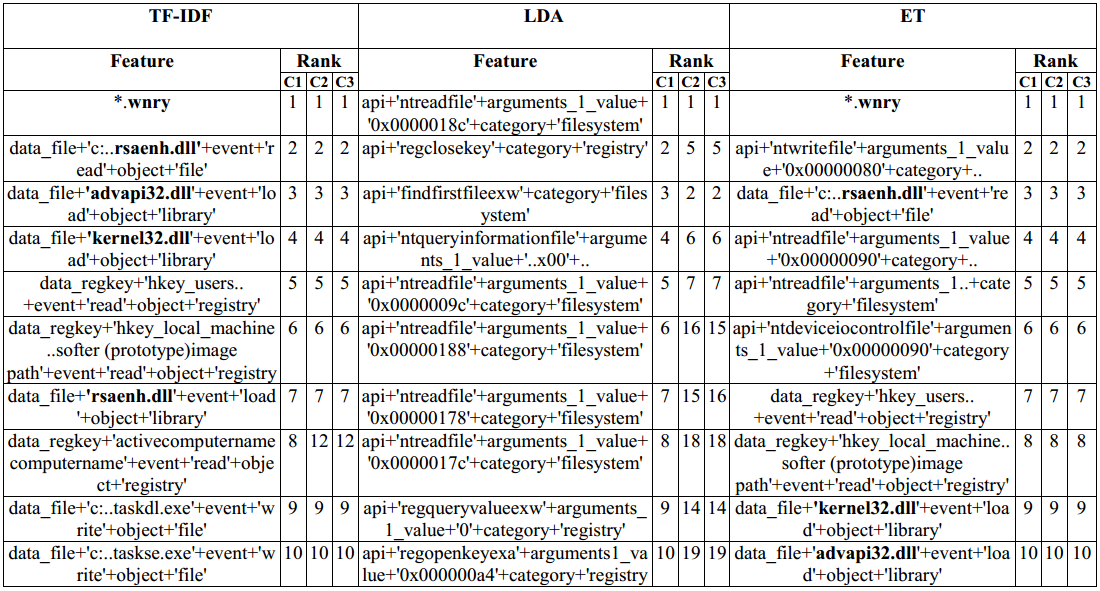}
\vspace{-0.2 cm}
\end{table}
 %=====
%\begin{wrapfigure}{r}{0.6\linewidth}
\begin{figure}[h]
\vspace{-1mm}
\centering\includegraphics[scale=0.7]{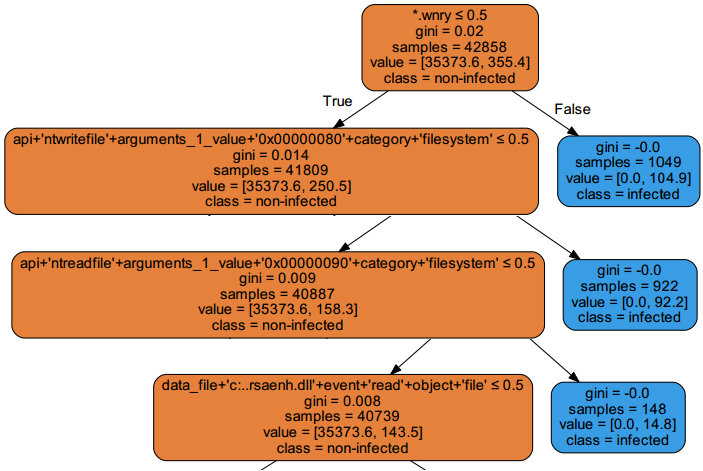}
\caption{Decision path based on the training logs showing how the most discriminating features are correlated in the decision making process. %The non-leaf brown boxes are features. A brown leaf node represent the non-infected class and a blue leaf node represent the infected class.
}
\label{fig:tree1}
\vspace{-1mm}
\end{figure}
%\end{wrapfigure}
%====

Table~\ref{table:experiment2} presents the top ten features of WannaCry that are calculated by the three machine learning methods when the ambient logging data are different. The experimental results present that the ET method is robust to provide the same rankings of the top ten features under the three tested scenarios. TF-IDF is less robust than ET, but Fisher's LDA provides completely different rankings of the top ten features in three different scenarios. Similar results were found when analyzing the top-ranked features of the other six ransomware attacks. Therefore, the ET algorithm is more robust to varying training data containing different quality and quantity of normal activity.\\ %Therefore, ET is a better ransomware detection model that is resilient to data distribution.   

\noindent \textbf{Experiment Three: Ransomware Early Detection}. 
The ET decision tree classifier is applied to detect the seven ransomware before encryption from a large majority of non-malicious activity. Table~\ref{table:experiment5} presents the detection rate of the seven ransomware attacks. Note that while recall varies, meaning the method produces false negatives, precision is always perfect, meaning there are no false positives. In terms of overall performance metrics, the detection model Gandcrab performs the best and DBGer performs the worst.  We also create graphs of each  decision tree to better interpret and visualize the detection results. Using  WannaCry attack as an example, Figure~\ref{fig:tree1} displays  first three levels of the decision tree. 
%======
%\begin{wraptable}[9]{r}{0.7\linewidth}
%\vspace{-.5cm}

\begin{table}
\centering
\caption{ET Early Detection Results}
\label{table:experiment5}
\begin{tabular}{|c|c|c|c|c|}
\hline
Ransomware              & Accuracy & Precision & Recall & F-Score \\ \hline
WannaCry                & 0.918    & 1         & 0.717  & 0.835   \\ \hline
\textit{\textbf{DBGer}} & 0.987    & 1         & 0.308  & 0.471   \\ \hline
Defray                  & 0.994    & 1         & 0.992  & 0.996   \\ \hline
Locky                   & 0.997    & 1         & 0.806  & 0.893   \\ \hline
Cerber                  & 0.987    & 1         & 0.505  & 0.671   \\ \hline
\uline{ \textbf{GandCrab}} & 0.999    & 1         & 0.997  & 0.999   \\ \hline
nRansom                 & 0.994    & 1         & 0.382  & 0.553   \\ \hline
\end{tabular}
\end{table}
 
The brown non-leaf nodes (rectangular boxes) represent the features of normal activity and the blue non-leaf nodes represent features induced by WannaCry. 
By retrieving the blue nodes on the top of the decision tree,  we can identify WannaCry's discriminating features. The correlation coefficients of these features are provided in non-leaf boxes. The graphs facilitate malware forensics analysis and allow operators to visualize  disruptive activity and determine the damages induced by the malware for proposing an optimal protection and response plan.

\section{Conclusion}
\label{sec:con}

We develop an automated ransomware pattern-extraction and early detection tool that extracts the sequence of events induced by seven ransomware attacks,  identifies the most discriminating features using three machine learning methods, and creates graphs to facilitate forensic efforts by visualizing features and their correlations. The experimental results present that TF-IDF feature ranking yields the most accurate identification of the ransomware-discriminating features, while the ET method is the most time efficient and robust to the variation of inputs. 
Notable, discriminating features are automatically promoted by this method that malware analysis reports failed to identify. 
%There are several unresolved problems for future research. First, this study uses Cuckoo Sandbox to provide detailed host behavior reports that contain various types of activity (e.g., file, registry, API calls). However, 

As the target application is using this to analyze real host logs collected by SOCs, future research to test our tool using real-world host-based data captured in enterprise networks to determine conditions for success. 
% to profile and build a detector of a new ransomware upon the initial infection. 
% Yet, %much fewer host-based events are monitored daily in some enterprises. The 
% configuration settings of the monitoring systems depend on organization's security goals and resources \cite{bridges2018information}.  %For example, only 20 Windows events are monitored and analyzed in the publicly available dataset from Los Alamos National Laboratory's enterprise network~\cite{WLS}.  
% Therefore, future research should test our tool using real-world host-based data captured in enterprise networks to determine conditions for success. % the efficiency of the pattern-extraction and early detection. 
Moreover, large enterprises generate large volumes of host data. The offline machine learning techniques used in this paper---creating features from host logs, determining malware discriminating features and detecting attacks---may not scale. 
Future research  using online machine learning technique (e.g., incremental decision tree) and deep learning methods (e.g., LSTMs) can enhance the tool.  

%Moreover, the classification of records into malicious or non-malicious also shows an accuracy between 91 and 99 percent for different individual ransomware. The generalized technique that can say whether a log is malicious or not also exhibits an accuracy rate of 98 percent. Though the generalized approach is not capable to identify the exact type of ransomware, which is a future direction of research.  
\section*{Acknowledgements}
Special thanks to the reviewers that helped polish this document, including Michael Iannacone. Research sponsored by the Laboratory Directed Research and Development Program of Oak Ridge National Laboratory, managed by UT-Battelle, LLC, for the U. S. Department of Energy, and by the National Science Foundation under Grant No.1812599. 
Any opinions,
findings, and conclusions or recommendations expressed in
this material are those of the authors and do not necessarily
reflect the views of the National Science Foundation.

\small
\bibliographystyle{IEEEtran}
\bibliography{main.bib}

\end{document}